\documentstyle[12pt]{article}

\def\@cite#1#2{$\m@th^{\the\scriptfont0 {#1\if@tempswa , #2\fi} }$}

\def\thebibliography#1{
\section{{\bf REFERENCES}}
\list{\arabic{enumi}.}
{\footnotesize
\settowidth\labelwidth{[#1]}
\addtolength{\labelwidth}{2pt}
\setlength{\itemsep}{0in}  \setlength{\parsep}{0in}
\setlength{\topsep}{0in}
\itemindent -0.11in
\leftmargin\labelwidth
\advance\leftmargin\labelsep
\usecounter{enumi}
}
 \def\newblock{\hskip .11em plus .33em minus .07em}
 \sloppy\clubpenalty4000\widowpenalty4000
 \sfcode`\.=1000\relax
}

\def\verse{\let\\=\@centercr
  \list{}{\itemsep\z@ \itemindent -1.5em\listparindent \itemindent
          \rightmargin\leftmargin\advance\leftmargin 1.5em}\item[]}

\input prepictex
\input pictex
\input postpictex

\begin{document}
\pagestyle{empty}
\setlength{\baselineskip}{0.1875in}
\renewcommand{\section}[1]{\vspace{\baselineskip}
\noindent {\bf #1} \vspace{\baselineskip}}
\def\thefootnote{\fnsymbol{footnote}}

\newcommand{\nc}{\newcommand}
\nc{\beq}{\begin{equation}}   \nc{\eeq}{\end{equation}}
\nc{\beqa}{\begin{eqnarray}}  \nc{\eeqa}{\end{eqnarray}}
\nc{\lsim}{\begin{array}{c}\,\sim\vspace{-21pt}\\< \end{array}}
\nc{\gsim}{\begin{array}{c}\sim\vspace{-21pt}\\> \end{array}}
\nc{\create}{\hat{a}^\dagger}   \nc{\destroy}{\hat{a}}
\nc{\kvec}{\vec{k}}             \nc{\kvecp}{\vec{k}^\prime}
\nc{\kvecpp}{\vec{k}^{\prime\prime} }   \nc{\kb}{\bf k}
\nc{\kbp}{{\bf k}^\prime}       \nc{\kbpp}{{\bf k}^{\prime\prime} }
\nc{\bfk}{{\bf k}}              \nc{\cohak}{a_{{\bf k}}}
\nc{\cohap}{a_{{\bf p}}}        \nc{\cohaq}{a_{{\bf q}}}
\nc{\cohbk}{b^*_{{\bf k}}}      \nc{\cohbp}{b^*_{{\bf p}}}
\nc{\cohbq}{b^*_{{\bf q}}}

\begin{flushleft}
{\bf QUANTUM SCATTERING
AND CLASSICAL SOLUTIONS\footnote{Talk presented at Sintra '94
workshop
on Electroweak Physics and the Early Universe.}}
\end{flushleft}
\vspace{\baselineskip}
\begin{tabbing}
\hspace{1.0in} \= Stephen D.H. Hsu \\[\baselineskip]
\> Lyman Laboratory of Physics\\
\> Harvard University\\
\> Cambridge, MA 02138
\end{tabbing}
\smallskip
{\bf Abstract}
\begin{quote}
I discuss a formalism for computing quantum scattering amplitudes
using a semiclassical expansion of a functional integral
representation
for the S-matrix.
The classical background for the expansion is determined by
solving the equations of motion subject to nontrivial boundary
conditions
determined by the initial and final quantum states.
The formalism
is designed to accomodate intrinsically nonperturbative processes
such
as baryon number violation, quantum tunneling and multiparticle
scattering.
It appears to
yield a controlled small coupling expansion even at asymptotically
high energies where instanton methods fail.

\end{quote}

\vspace{\baselineskip}

\section{INTRODUCTION}

The behavior of quantum scattering amplitudes at large center of mass
energies,
$E \sim E_*  \equiv m/g^2$,
is an open question currently under investigation.
Here,
$m$ is a mass scale characterizing the particle excitations of the
theory,
and $g$ is a small coupling constant controlling the nonlinear
interactions.
It has been speculated that multiparticle scattering from initial
states
with only a small number of particles could be unsuppressed
at these energies.
A second related issue is whether anomalous processes,
such as baryon number violation in electroweak theory where
$E_*\simeq {\cal O}\left(M_w/\alpha_w\right)\sim 10~TeV$,
are unsuppressed at energies of the same order~\cite{BNV}.

Standard perturbation theory is clearly inadequate to address either
of
these issues.
Recent investigations have focused on expansions about
nonperturbative
solutions of classical field equations, such as
instantons~\cite{BNV,TIN}.
However,
none of the previously proposed solutions provides a stationary point
for a controlled semiclassical expansion at the energies of interest,
$E\sim E_*$~\cite{MMY}.
In general,
the semiclassical expansion about the proposed instanton-like
saddlepoints
is controlled only at low energies and provides little or no
information at $E \sim E_*$.

In this talk,
I will discuss a method for finding classical solutions
which will prove suitable saddlepoints of scattering amplitudes
even at $E \sim E_*$.
The equations and boundary conditions which must be satisfied are
dramatically simplified compared to a previous approach
\setlength{\topskip}{0.0in}
to the same problem by Mattis, McLerran and Yaffe~\cite{MMY}.
In particular,
the resulting equations and boundary conditions for the classical
saddlepoint are no longer of the integro-differential type.
Rather,
the equations of motion are source-free,
$$
{\delta S \over \delta \phi(x)} = 0
$$
and in certain cases the boundary conditions are of the initial value
type,
which can be integrated forward in time using standard numerical
algorithms.
This is also a departure from the approach of Khlebnikov, Rubakov and
Tinyakov~\cite{TIN,KRT}, who derive
similar source-free equations, but for an initial {\em coherent}
state,
rather than a wave-packet type state which is relevant for high
energy
collisions.

\setlength{\topskip}{0.0in}
In light of this new formalism,
I will discuss the implications of some recent numerical simulations
in
classical
field theory for high energy two-particle scattering.
In particular, in 4-dimensional pure SU(2) Yang-Mills theory,
Gong, Matinyan, M\"uller and Trayanov~\cite{GMMT} have found an
instability
of high frequency standing waves of pure glue to decay to low
frequency modes.
When interpreted within our formalism,
this result leads to the speculation that multiparticle scattering in
$SU(2)$ gauge theories may be unsuppressed (at least exponentially)
at
high energies.
A generalization of our method to complex trajectories
which allows their application
to essentially any wave packet - coherent state scattering amplitude.
I finally discuss how our formulation relates to some previous work
by other authors on high energy multiparticle and anomalous processes
in electroweak theory~\cite{MMY,KRT}.

Before launching into the formalism, let me make a few remarks on the
method and the results we can expect.
We are interested in quantum scattering amplitudes (or S-matrix
elements)
between arbitrary initial and final states $\vert i \rangle$ and
$\vert f
\rangle$.
In most of what follows in the talk I will be specifically interested
in states
that are
relevant to accelerator experiments (i.e. wave packet states), but
here I will
be
completely general.

Now, imagine thinking about the scattering process in terms of paths
inside the
functional integral. The quantum fields do whatever they want,
weighted by the
appropriate action $exp[ i S ]$, but also by the overlap of the
asymptotic part
of
the path with the initial and final states. Therefore we expect that
the
amplitude must
be expressible in the form:
\beq
\label{heur}
\langle f \vert S \vert i \rangle ~\sim~
\int d \phi_i~ d \phi_f ~D \phi~ \Psi_i [ \phi(T_i)] ~ \Psi_f
[\phi(T_f)]~ e^{i
S[\phi]},
\eeq
where I have explicitly written the measure of path integration to
include
fluctuations
of the fields in the asymptotic past and future. The wave-functionals
$\Psi_{i,f}$
measure the overlap of the initial and final states with position (or
field
operator)
eigenstates $\phi$ at asymptotic times. Below I will give an actual
derivation
of
(\ref{heur}) along with the explicit form of the wavefunctionals in
the case
that the initial
and final states are either wave-packets or coherent states. (Recall
that a
plane wave is
a special type of  wave packet that is narrowly peaked in momentum
space, so we
can
work with plane waves as well if we wish.)

Now, the usual way to compute (\ref{heur}) at weak coupling is by
expanding the
functional integral
about either the vacuum - which generates standard perturbation
theory - or
about
a more exotic classical path such as the instanton, but which also
happens to
obey
vacuum boundary conditions.  (In the case of the instanton
(\ref{heur}) is
first continued
to Euclidean space, where finite action solutions must satisfy vacuum
boundary
conditions.
The results of expanding the path integral are then analytically
continued back
term by term.
More on this later.)
Clearly, for some initial and final states it is not a big
loss to evaluate the overlaps $\Psi_{i,f} [ \phi ]$ on the asymptotic
vacuum.
We
know that in most instances standard perturbation theory works quite
well.
We also know that instanton methods give a controlled way of
computing
tunneling
phenomena - at least at low energies.

However, to get a better approximation to the above path integral,
one could
try to
extremize the entire integrand, which will yield some nontrivial
boundary
conditions
on the classical solutions, as determined by the specific form of the
wavefunctionals,
or in other words the initial and final states of interest.

\section{FORMALISM}

As I mentioned above,
we wish to calculate
the \mbox{S-matrix} element between an initial two particle state
and an arbitrary final state~\footnote{The authors of \cite{MMY}
extremize
momentum-space Greens functions. See Section 6.}.
The result  will be a boundary value problem which
determines the classical field which is the stationary point of the
\mbox{S-matrix} element.

To this end,
we express the kernel of the \mbox{S-matrix} in a basis of coherent
states, first used in this context in \cite{KRT}.
The initial and final states are defined by sets of complex variables
${\bf a}\equiv \{\cohak\}$, ${\bf b}^*\equiv \{\cohbk\}$,
respectively.
A coherent state $\vert \cohak \rangle$ is an eigenstate of the
annihilation
operator $\destroy_{\bf k}$:
$\destroy_{\bf k} \vert \cohak \rangle = \cohak  \vert \cohak
\rangle$.
The relevance of coherent states for the scattering problem stems
from
the fact that upon differentiation they generate all momentum
eigenstates (see eq.\ref{f2} below).

For simplicity,
we illustrate the case of a single real massive scalar field
with self-coupling $g^2$ in $3\! +\! 1$ dimensions.
The results can be generalized to the case of gauge fields.
First,
we express the transition amplitude from an initial coherent state
$\vert {\bf a}\rangle$ at time $T_i $ to a final coherent state
$\vert {\bf b}^*\rangle $ at time $T_f $,
in terms of the path integral by inserting a complete
set of eigenstates of the field operator on the initial and final
time
slices (see figure 1)
\beq
\label{transamp}
\langle\, {\bf b}^*\,\vert \, U\,\vert\, {\bf a}\,\rangle ~=~
\int d\phi_i\, d\phi_f ~\langle\, {\bf b}^*\,\vert\,\phi_f\,\rangle~
\langle\,\phi_f\,\vert\, U\,\vert\,\phi_i\,\rangle~
\langle\,\phi_i\,\vert\, {\bf a}\,\rangle ~.
\eeq

$$
\beginpicture
\setplotarea x from -65 to 65 , y from -40 to 55
\ellipticalarc axes ratio 2:1 180 degrees from -40 0 center at 0 0
\ellipticalarc axes ratio 2:1 360 degrees from 40 80 center at 0 80
\putrule from 40 0 to 40 80    \putrule from -40 0 to -40 80
\arrow < 4 pt> [0.4,1.5] from 0 82 to 0 110  \put{$t$} at 0 120
\putrule from 45 0 to 55 0     \putrule from 45 80 to 55 80
\put{$T_i$} at 65 0            \put{$T_f$} at 65 80
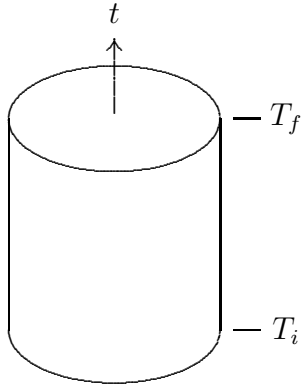
\put{ Figure 1: Spacetime, with two slices.  } at 200 -20
\endpicture
$$

$U$ is the evolution operator between time $T_i $ and $T_f $.
A``position'' eigenstate of the field operator $\phi $ is denoted
$\vert\,\phi\,\rangle $ and $\phi_{i,f}=\phi(T_{i,f}) $.
Then,
from (\ref{transamp}), we obtain
the \mbox{S-matrix} kernel in a compact form in terms of path
integrals
\beq
\label{s}
\langle\, {\bf b}^*\,\vert\, S\,\vert\, {\bf a}\,\rangle ~\equiv~
S\left[ {\bf b}^*, {\bf a} \right] ~=~
\lim_{T_i,T_f \rightarrow \mp \infty}~
\int d\phi_f\, d\phi_i~ e^{B_f}~e^{B_i}
\int_{\phi_i}^{\phi_f} D\phi~ e^{iS\,\left[\,\phi\,\right]} ~,
\eeq
where $S\left[\phi\right]$ is the action functional.
The path integral appearing here is over fields obeying the boundary
conditions $\phi(T_{i,f}) = \phi_{i,f}$.
The functional $B_f$ is
\beqa
\label{bf}
B_f\left[ {\bf b}^*, \phi_f\right] & = &
- {1\over 2} \int\! d^3k ~ \cohbk b^*_{\bf -k} ~e^{2i\omega_kT_f}
- {1\over 2} \int\! d^3k ~\omega_{\bf k} ~\phi_f( \kvec )
{}~\phi_f(-\kvec ) \\
&  & +~\int\! d^3k ~\sqrt{2\omega_{\bf k}}~e^{i\omega_k T_f}~
\cohbk~\phi_f(-\kvec ) \nonumber ~,
\eeqa
in terms of which the wave functional of the final coherent state is
\beq
\label{finalwf}
\langle\, {\bf b}^*\,\vert\,\phi_f\, \rangle ~\equiv~
\exp{\left(\, B_f\left[ {\bf b}^*,\phi_f\right] \,\right)} ~.
\eeq
Similarly, the functional $B_i$ is
\beqa
\label{eq:bi}
B_i \left[ {\bf a}, \phi_i \right] &  = &
- {1\over 2} \int d^3k ~\cohak a_{\bf -k} e^{-2i\omega_k T_i}
- {1\over 2} \int d^3k ~\omega_{\bf k} \phi_i(\kvec ) \phi_i(-\kvec )
\\
&  & +~
\int\! d^3k ~\sqrt{2\omega_{\bf k}}~e^{-i\omega_k T_i}~
\cohak~\phi_i(\kvec ) \nonumber
\eeqa
is related to the wave functional of the initial coherent state
\beq
\label{initialwf}
\langle\,\phi_i\,\vert\, {\bf a}\,\rangle  ~\equiv~
\exp{\left(\, B_i\left[ {\bf a},\phi_i\right] \,\right)} ~.
\eeq

Here the 3-dimensional Fourier transform is defined
\beq
\label{FT}
\phi_{i,f}( \kvec ) ~=~
\int {d^3x \over (2\pi)^{3/2} }~
e^{i\vec{k}\cdot\vec{x}} ~\phi (T_{i,f}, \vec{x}) ~,
\eeq
and is related to the residue of the 4-dimensional Fourier
transform when the field reduces to a plane wave superposition
as $\vert\, T_{i,f}\,\vert\rightarrow\infty $~\cite{KRT}.

The kernel (\ref{s}) is a generating functional for \mbox{S-matrix}
elements between any initial and final $N$ particle states,
by functional differentiation with respect to arbitrary $\cohak$ and
$\cohbk$.
We now use this fact to construct a kernel for scattering from
initial
two particle states.
We define an initial two particle (wave packet) state at $t = T_i$
\beq
\label{alpha}
\vert\,\vec{p}, -\vec{p}\,\rangle ~\equiv~
\int d^3k ~\alpha_R(\kvec ) ~\create_{\bf k}
\int d^3k^\prime ~\alpha_L(\kvecp)~\create_{\kbp}~
\vert\, 0\, \rangle ~,
\eeq
where $\create_{\kb} $ is a creation operator,
and $\alpha_{R,L}(\kvec )$ are arbitrary smearing functions
of $\kvec $,
localized around some reference momenta
$\vec{p}$ and  $-\vec{p}$ respectively.
The wave packets are
normalized so that~\footnote{We use the
non-relativistically normalized commutator
$\left[\,\destroy_{\kbp}\, ,\, \create_{\kb} \,\right] ~=~
\delta^3\left( \kvec - \kvecp \right) $.}
\beq
\label{alphanorm}
\int d^3k ~\vert\, \alpha_{R,L}(k )\, \vert^2 ~=~ 1  ~.
\eeq
The regime relevent to the high energy multiparticle scattering
problem
is where, in the limit $g^2\rightarrow 0$,
the wave packet is localized around momenta
$|p|\sim m/g^2$ with a characteristic width, $\Delta p \sim m$.

This state can be generated by functional differentiation
of the coherent state $\vert {\bf a}\rangle $ with
respect to $ a_{\bf k} $
\beq
\label{f2}
\vert\vec{p}, -\vec{p}\rangle ~=~
\int d^3\kvec~d^3\kvecp~
\alpha_R(\kvec ) ~\alpha_L(\kvecp)~
{\delta\over\delta a_{\bf k}}~{\delta\over\delta a_{\kbp}}~
\vert\, {\bf a}\,\rangle ~\rule[-3mm]{0.2mm}{9mm}_{~{\bf a}\, =\, 0}
{}~.
\eeq

So, differentiating under the functional integral,
the \mbox{S-matrix} element between the two particle state (\ref{f2})
and any final state $\vert\{\cohbk\}\rangle$ involves the following
functional at $t = T_i$
\beqa
\label{bi2}
\lefteqn{
{\delta\over\delta a_{\bf k}} {\delta\over\delta a_{\kbp}}
\exp{\left(\, B_i\left[{\bf a},\phi_i\right] \,\right)}~
\rule[-3mm]{0.2mm}{9mm}_{~a = 0} ~ =} & & \\
& & \hspace{2cm}
2~\sqrt{\omega_{\bf k}\omega_{\kbp}}~
\phi_i(\kvec)~\phi_i(\kvecp)~e^{-i(\omega_{\bf k}+\omega_{\kbp})T_i}
\exp{\left(\, B_i\left[0, \phi_i\right] \,\right)}\nonumber ~,
\eeqa
after dropping a term which vanishes in the limit $T_i \rightarrow
-\infty$.
The last factor here is simply the normalization of the initial
position eigenstate
\beq
\exp{\left(\,
-{1\over 2}\int d^3k~\omega_k~\phi_i(\vec{k} )~\phi_i(-\vec{k} )
\,\right)} ~.
\eeq

We combine this with the smearing functions and finally
obtain an \mbox{S-matrix} kernel for the scattering of two wave
packets
into arbitrary final states,
\beq
\label{combine}
S\left[{\bf b}^*, 2 \right] ~=~
\lim_{T_i,T_f\rightarrow \mp\infty}
\int d\phi_f\, d\phi_i~
\alpha_R\cdot\phi_i~\alpha_L\cdot\phi_i ~
e^{B_f\left[ b,\phi_f\right] ~+~ B_i\left[0, \phi_i\right] }
\int_{\phi_i}^{\phi_f} D\phi~e^{iS\,\left[\,\phi\,\right]}~,
\eeq
where we have denoted the initial state (\ref{f2}) by ``$2$''.
Here we have used the following compact notation for
the initial state factors
\beq
\label{integral}
\alpha\cdot\phi_i ~\equiv~
\int d^3k ~\sqrt{2\omega_k}~
\alpha(k ) ~\phi_i(k ) ~e^{-i\omega_k T_i} ~.
\eeq

We now derive the boundary value problem obeyed by
the classical field in the stationary phase approximation
of (\ref{combine}).
The stationary phase approximation of ordinary integrals
suggests that the initial state factors  be included
in the extremization of the path integral~\footnote{The Stirling
approximation to the Gamma function is a classic example.}.
These factors will certainly impact the stationary phase
in the kinematic regime relevant to the high energy scattering
problem, where the initial wave packets are peaked around
$\vert p\vert\sim m/g^2$, as expansions around the instanton
indicate~\cite{KRT}.
More generally though, we will show in Section 4 that this
procedure provides a consistent weak coupling expansion at {\it any}
fixed center of mass energy.

We include initial state factors in a straightforward manner by
first exponentiating the initial state factors into an
``effective action'', so that
\beq
\label{seff}
S\left[ {\bf b}^*, 2\right] ~=~
\lim_{T_i,T_f\rightarrow\mp\infty}
\int d\phi_f\, d\phi_i\, D\phi ~e^{\Gamma}  ~,
\eeq
where the  effective action $\Gamma$ is
\beq
\label{gam}
\Gamma\,\left[\,\phi\,\right] ~=~
\ln \alpha_R\cdot\phi_i~\alpha_L\cdot\phi_i ~+~
B_i\left[0, \phi_i\right]~+~
iS\,\left[\,\phi\,\right] ~+~ B_f\left[b^* , \phi_f\right] ~,
\eeq
after dropping a term which vanishes as $T_i\rightarrow -\infty$.

We can now derive the boundary value problem by varying the effective
action.
Varying the entire exponent $\Gamma$
with respect to $\phi(x)$ for $T_i < t < T_f$
gives the  source-free equations of motion
\beq
\label{eom}
{\delta S \over \delta \phi(x)} ~=~ 0 ~.
\eeq
Varying the entire exponent with respect to $\phi_i(k)$, gives
\beq
\label{ti}
i\,\dot{\phi}_i (\vec{k}) ~+~ \omega_k \phi_i (\vec{k}) ~=~
\sqrt{2\,\omega_{\bf k}}~\left(\,
{\alpha_R(\vec{k})\over\alpha_R\cdot\phi_i}~+~
{\alpha_L(\vec{k})\over\alpha_L\cdot\phi_i}\,\right)
{}~e^{-i\omega_k T_i} ~.
\eeq
The first term on the left hand side comes from a surface term in
the action $S$.
The other terms come from variation of the wave functional at
$t=T_i$.
This boundary condition involves both the positive and negative
frequency parts of the field,
unlike the boundary condition which arises for an initial coherent
state $\vert{\bf a}\rangle$,
which depends on the negative frequency component of $\phi$
only~\cite{TIN}.

The boundary condition (\ref{ti}) at the initial time slice is rather
complicated.
However, it can be simplified since a real field $\phi$ may be
written
in the asymptotic region $t = T_i\rightarrow -\infty$ as a plane wave
superposition
\beq
\label{phialpha}
\phi_i(\vec{k}) ~=~ {1\over \sqrt{2\,\omega_{\bf k}}}~
\left(\,
u_{\bf k}~e^{-i\omega_k T_i} ~+~ u_{-\bf k}^*~e^{i\omega_k T_i}
\,\right) ~.
\eeq
Equation (\ref{ti}) then reduces to the requirement
\beq
\label{u}
u_{\bf k} ~=~
{
\alpha_R(\vec{k} ) ~+~ \alpha_L(\vec{k}) \over
\left(\,
1 ~+~ \int d^3k ~\alpha_R(\vec{k} )~\alpha_L(\vec{k})
\,\right)^{1/2}
}~,
\eeq
using the normalization (\ref{alphanorm}).
This solution is consistent with physical intuition,
the classical field reducing to the initial particles at early times.
The overlap of the left- and right- moving wave packets in the
denominator is very small for narrow high energy wave packets.

If the field is real,
then the negative frequency part
equals the complex conjugate of the positive frequency part,
and the field $\phi_i$ is determined as in (\ref{phialpha}) and
(\ref{u}).
The real initial condition can be integrated forward, and
uniquely determines the final field $\phi_f$.
We say more about the case of {\it complex} stationary points of
real fields in Section 5.

It remains to be seen in what sense it is a good approximation
to include the initial state prefactors in the stationary phase
calculation.
We will demonstrate in Section 4 that the stationary phase solution
which results from this approach provides a {\it controlled} weak
coupling expansion of the scattering amplitude.

Finally, varying with respect to $\phi_f(k)$ gives
\beq
\label{tf}
-i\,\dot{\phi}_f(k) ~+~ \omega_{\bf k}\phi_f(k) ~=~
\sqrt{2\,\omega_k}~ b^*_{-\bf k} ~e^{i\omega_k T_f }~.
\eeq
Again with a free field form as in (\ref{phialpha}),
this boundary condition may be re-expressed as
\beq
\label{tf2}
\phi_f(\vec{k}) ~=~ {1\over \sqrt{2\,\omega_{\bf k}}}~
\left(\,
b_{\bf k}~
e^{-i\omega_k T_f} ~+~  b_{-\bf k}^*~e^{i\omega_k T_f}
\,\right) ~.
\eeq
We see that the precise scattering amplitude which is extremized is
only
determined at the end of the calculation,
by the asymptotic form of the classical solution in the far future,
$T_f \rightarrow \infty$.
For the initial conditions given by (\ref{ti}),
the scattering amplitude corresponds to a transition from wave packet
states to a final coherent state described by $\vert b^*\rangle$
from (\ref{tf}).
Note that neither a wave packet state~\footnote{Except in the limit
of
$\alpha_R(\kvec ) \sim \delta^3 (\kvec - \vec{p})$,
where it reduces to a plane wave.} nor a coherent state are
eigenstates
of the Hamiltonian, so that neither have a definite energy.

When the classical field satisfies the equation of motion (\ref{eom})
and the boundary conditions (\ref{ti}) and (\ref{tf}),
there are no initial and final state corrections to leading order
in the semiclassical expansion.
This is not the case for the instanton, which satisfies the
equation of motion but not the correct boundary conditions.
Instead,
there are linear terms in the fluctuation about the instanton
and these generate initial and final state corrections.

\section{SEARCH FOR SOLUTIONS}

The results of the previous section establish an important fact:
\begin{quote}
The classical field obtained by starting with two incident wave
packets
and evolving them forward using (\ref{eom}) is the dominant
contribution
to {\it some} two particle scattering amplitude, in the semiclassical
approximation.
\end{quote}
In the case of a purely Minkowskian classical solution,
the result for the corresponding scattering amplitude will {\it not}
contain an exponential suppression.
Thus,
if a real time classical trajectory can be found which connects
wave packet initial conditions to an ``interesting''
final state, for example a multiparticle or fermion number violating
state,
our results imply that the corresponding scattering amplitude is
unsuppressed.
Classical solutions which either exist in complex time or are
themselves
complex are relevant to classically forbidden transitions and will be
discussed in \mbox{Section 5.}

The boundary conditions given by (\ref{phialpha}) and (\ref{u}) are
amenable
to straightforward numerical integration.
The boundary conditions specify $\phi(x,T_i)$ and $\dot{\phi}
(x,T_i)$, which
are sufficient to construct $\phi(x,t)$ for all subsequent $t > T_i$,
given a discretization of the equations of motion (\ref{eom}).
In principle,
it is possible to generate an infinite number of classical
trajectories,
each relevant to a particular initial wave packet scattering
amplitude.

Of course,
it is not clear {\it a priori} that the final state which results
from the classical evolution will be one which is interesting to the
problem of multiparticle production or anomalous baryon number
violation.
However,
the above observation summarizes the relevance for scattering
amplitudes
of computations by Rajagopal and Turok~\cite{RT}, and also Goldberg,
Nash
and Vaughn~\cite{GNV}.
Rajagopal and Turok studied the classical scattering of wave packets
in the
Abelian Higgs model,
while Goldberg et al. studied the classical dynamics of $\phi^4$
theory.
Neither group found that energy was readily transferred from high
frequency
to low frequency modes,
a signal which would indicate the production of a final state with
many
(soft) particles.
Therefore,
the scattering amplitudes which are dominated by their classical
trajectories are probably not relevant to multiparticle production
or fermion number violation.

On the other hand, the initial indications on the behavior of
non-Abelian
classical trajectories seem more promising.
Some recent results on the classical behavior of pure $SU(2)$
Yang-Mills
theory by Gong, Matinyan, M\"uller and Trayanov~\cite{GMMT} may
provide insight
into solutions of the boundary value problem presented above.
These authors have considered the classical stability of a stationary
mono-color wave in Yang-Mills theory
\mbox{(in $A^c_0 = 0$ gauge):}
\beq
\label{stwv}
A^c_i (x,t) ~=~ \delta_{i3} ~\delta_{c3} ~A ~\cos k_0x ~\cos \omega_0
t,
\eeq
where $c$ is a color and $i$ is a spatial
index \footnote{NB - A {\it travelling} wave
is stable in Yang-Mills theory~\cite{travelling}.}.
Small amplitude variations of the field in directions of different
color
are found to lead to an instablility with long wavelength.

The Hamiltonian
evolution of a perturbed standing wave is equivalent to the evolution
of the initial conditions relevant to the scattering of two perturbed
plane
waves in  $SU(2)$ gauge theory.
The discovery of an instability of the initial
configuration to decay into long wavelength modes implies the
existence of
a classical trajectory connecting initial high energy plane waves to
long
wavelength modes in the final state.
It therefore suggests that the corresponding $2 \rightarrow$ {\it
many}
gluon scattering amplitude may be unsuppressed !
The relevance of weak coupling calculations to
pure gauge theory is unclear, due to the asymptotic freedom of the
theory.
However, it is possible that arguments similar to those from deep
inelastic scattering
may be used to justify the semiclassical result.
In the limit that all of the energy scales (including the energies of
the
many ``soft'' outgoing gluons) are large compared to the intrinsic
mass scale
of the theory $\Lambda_{SU(2)}$,
it seems plausible that the relevant running coupling constant is
small.

It is also important to determine whether the instability persists
for
wave packet initial conditions, as the plane wave limit is never
achieved at an actual accelerator. Since wave packets are
localized in space, two packets have only a finite amount of time
to interact before passing completely through each other. This is
in contrast to plane waves, which exist everywhere in spacetime.
Since the plane waves of M\"uller et al.~\cite{GMMT} have finite,
nonzero
amplitude,
they also contain the equivalent of an infinite number of particles.
It is
quite possible that the behavior of wave packets representing a
finite number of particles is very different.

Another class of trajectories in pure $SU(2)$ Yang-Mills theory are
those
found by Farhi et al.~\cite{Farhi}.
These solutions in the spherical ansatz  correspond to spherical
shells of glue that originate at infinity, collapse inward, and
bounce
back to infinity, leaving behind some fractional topological charge.
There are some open questions concerning such solutions - in
particular,
their stability to perturbations which are outside the spherical
ansatz,
and their relevance to fermion number violation.
However, in the absence of fermions, they can be interpreted within
our
formalism as extremizing scattering amplitudes between spherically
symmetric
initial and final quantum states.

Of course, our eventual goal is to address the electroweak theory,
where the Higgs mechanism cuts off the infrared growth of the
coupling
constant.
It would be extremely interesting if simulations of wave packet
scattering
could be performed in a spontaneously broken $SU(2)$ gauge theory.

Clearly, there is much numerical work to be done to
address these issues.

\section{THE EXPANSION}

Let us now
construct the perturbative expansion around the classical
solution of the BVP.
We will demonstrate that corrections to the
leading contribution are systematically suppressed in
powers of the small coupling constant for any fixed energy.
The resulting small coupling expansion differs from the
usual one (e.g.-- instanton) and the subtleties involved
will be discussed at the end of this section.

As usual,
we proceed by expanding the quantum field $\phi$ around
a classical background field $\phi^c $, which is a solution
to the classical boundary value problem derived in Section 2.
\beq
\label{expand}
\phi(x) ~=~ \phi^c(x; \mu) ~+~ \nu(x) ~.
\eeq
The classical solution depends in general on a collection
of collective coordinates denoted here by $\mu$,
corresponding to the invariances of the
{\em scattering amplitude}~\cite{MMY,GH}.
We expect invariances of translation and scale size to be broken
by the choice of $\alpha_{R,L}(\kvec )$; however,
$\mu$ may include gauge coordinates if we are working in a gauge
theory.
Equation (\ref{expand}) involves an expansion on the initial and
final time
slices,
which we can express in momentum space as
\beq
\label{expandslice}
\phi_{i,f}(\vec{k}) ~=~
\phi^c_{i,f}(\vec{k}; \mu) ~+~ \nu_{i,f}(\vec{k}) ~~{\rm for}~~
t = T_{i,f} ~,
\eeq
where $\phi^c_{i,f}(\vec{k})$ is known explicitly for any given
initial and final wave packets, from (\ref{phialpha}) and (\ref{tf2})

Note that the classical solution is of order ${\cal O}(g^0)$,
since it is non-trivial  in the limit as $g\rightarrow 0$.
In fact, when $g$ vanishes,
the background field is simply the initial condition
propagated forward in time with the free Hamiltonian,
\beq
\label{phialphat}
\phi^c(t, \vec{k}) ~=~ {1\over \sqrt{2\omega_{\bf k}}}~
\left(\,
u_{\bf k}~
e^{-i\omega_k t} ~+~  u_{-\bf k}^*~e^{i\omega_k t}
\,\right) ~,
\eeq
where $u_{\bf k}$ is as in (\ref{u}).
In this limit, the two wave packets simply pass through each other.
For non-vanishing $g$,
the classical field $\phi^c$ will have some complicated
dependence on $g$, reflecting the non-linearity of the theory
and the external boundary conditions.

The kernel of the S-matrix (\ref{combine})
is composed of three functional integrals which we write as
\beq
\label{s2}
S\left[ {\bf b}^*, 2 \right] ~=~
\int d\phi_f~ d\phi_i~
\alpha_R\cdot\phi_i ~\alpha_L\cdot\phi_i~
e^{B_f[b^*,\phi_f]~+~B_i[0,\phi_i]}
\int_{\phi_i}^{\phi_f} D\phi~ e^{iS\,\left[\,\phi\,\right]}
\eeq
where the limit $|T_{i,f}|\rightarrow\infty$ is understood.
Now we formally expand each of the terms in (\ref{s2}) using
(\ref{expand}).
The action becomes
\beq
\label{sexpand}
S\left[\,\phi^c + \nu\,\right] ~ = ~ S\left[\,\phi^c\,\right] ~+~
\int
d^3x~\nu(x)~\dot{\phi}^c(x)~\rule[-3mm]{0.2mm}{8mm}_{\,T_i}^{\,T_f}~+
{}~
S_2\left[\,\nu\,\right] ~+~ S_{int}\left[\,\nu\,\right] ~,
\eeq
after integrating by parts and retaining the surface terms in the
time
direction, and using the equations of motion (\ref{eom}).
All of the non-quadratic dependence is contained in the interaction
terms $S_{int}\simeq {\cal O}(\nu^3, \nu^4)$ which are suppressed by
powers
of $g^2$.
Here $S_2$ is the part of the action quadratic in the fluctuation
field,
which can be expressed as
\beq
\label{action2}
S_2\left[\,\nu\,\right] ~=~
-\frac{1}{2}\int d^3x~\nu(x)~\dot{\nu}(x)~
\rule[-3mm]{0.2mm}{8mm}_{\,T_i}^{\,T_f}  ~-~
\frac{1}{2}\int d^4x~\nu(x)~\Delta\left[x\right]~\nu(x)~,
\eeq
in terms of the operator $\Delta $
of quadratic fluctuations in the $\phi^c$ background,
\beq
\label{quadop}
\delta^4(x - y)~\Delta\left[\, x\,\right] ~\equiv~
{\delta S\over \delta\phi (x)~\delta\phi (y)}~
\rule[-3mm]{0.2mm}{7mm}_{\,\phi^c} ~=~
\delta^4(x - y)
\left\{\,
\partial_\mu^2 ~+~ V^{\prime\prime}(\phi^c)
\,\right\} ~.
\eeq

Expanding the boundary functionals gives
\beq
\label{bfex}
B_f\left[\,{\bf b}^*, \phi^c_f + \nu_f \,\right] ~=~
-i\int d^3k ~\nu_f(k)~\dot{\phi}^c_f(-k) ~+~
B_f\left[\, {\bf b}^*, \phi^c_f\,\right] ~+~
B_f\left[\, 0, \nu_f\,\right] ~,
\eeq
after using the boundary condition (\ref{tf}), and
\beqa
\label{biex}
\lefteqn{B_i\left[\, 0, \phi^c_i + \nu_i \,\right] ~=~} & & \\
& & i\int d^3k~\nu_i(k)~\dot{\phi}^c_i(-k) ~+~
B_i\left[\, 0, \phi^c_i\,\right] ~+~
B_i\left[\, 0, \nu_i\,\right] ~-~
{\alpha_R\cdot\nu_i\over \alpha_R\cdot\phi^c_i} ~-~
{\alpha_L\cdot\nu_i \over \alpha_L\cdot\phi^c_i} \nonumber ~,
\eeqa
after using the boundary condition (\ref{ti}).
The $\dot{\phi}_{i,f}$ terms in (\ref{bfex}) and (\ref{biex}) cancel
with the boundary terms in the action (\ref{sexpand}).
Note the appearance of additional terms linear in $\nu_i$ in
(\ref{biex}),
which arise because we are expanding around a stationary point of
the {\em effective} action (\ref{gam}), instead of the action.
Note also that the terms independent of $\nu_{i,f}$ are the overlaps
of
the classical field with the initial and final states, respectively.

Now, substituting (\ref{sexpand}), (\ref{bfex}) and (\ref{biex}) in
(\ref{s2}), we obtain
\beq
\label{s2ex}
S\left[ {\bf b}^*, 2 \right] ~=~ \int d\mu~
e^{B_f[{\bf b}^*, \phi^c_f ] ~+~B_i[0,\phi^c_i] ~+~
iS\,\left[\,\phi^c\,\right]}~{\cal Z}\left[\,\phi^c\,\right]~
\left[\, 1 + {\cal O}(g^2)\,\right]
\eeq
where the quadratic integral over fluctuations is contained in
\beqa
\label{fancyz}
\lefteqn{
{\cal Z}\left[\phi^c\right] ~\equiv~} & & \\
& &
\int d\nu_f~ d\nu_i~ D\nu ~
\left\{\, 1 ~+~ \alpha_R\cdot\nu_i \,\right\}~
\left\{\, 1 ~+~ \alpha_L\cdot\nu_i\,\right\}~
e^{B_f[0, \nu_f] \, +\, B_i[0,\nu_i] \, -\,
\alpha_R\cdot\nu_i \, -\, \alpha_L\cdot\nu_i \, +\,
iS_2\,\left[\,\nu\,\right]} ~, \nonumber
\eeqa
after using $\alpha_{R,L}\cdot\phi^c_i = 1$ in the initial state
factors and in the exponent.
An integration over collective coordinates has been factored out
of (\ref{s2ex}) and
the remaining functional integral (\ref{fancyz}) involves only
fluctuations orthogonal to any zero modes of the classical
background.

The ${\cal O} (g^2)$ corrections result from the interactions in
$S_{int}$.
Expanding $\exp ( iS_{int} [ \nu ] )$ inside (\ref{s2}) yields the
leading
term plus corrections which have the form of vacuum to vacuum loops.
The vertices used to construct these loops each carry a suppression
of $g^2$.
The propagators inside these loops are the usual Feynman propagators
evaluated in the classical background $\phi^c$ ~\footnote{It is worth
remembering that since we are working directly in Minkowski space
(instead of beginning in Euclidean space and analytically continuing
back),
the Feynman boundary conditions on Greens functions are only obtained
through
some form of regularization, such as an $i \epsilon$ prescription.}.
In the instanton case~\cite{TIN,IS},
dangerous initial state corrections arose from the residues of
propagators
in the instanton background,
which displayed an energy dependence $\sim g^2 s$,
where $s$ is the center of mass energy of the collision and could be
parametrically of order $1/g^4$.
Note that it is only the {\em residue} of the instanton propagator
that
is ill-behaved and not the propagator itself.
Since the corrections in our case are proportional to loop integrals
rather than the residues of individual propagators \cite{TIN,IS},
we expect them to remain subleading even at high energies.

The Gaussian path integral appearing in (\ref{fancyz}) can now
be done exactly.
First factor out the quadratic prefactor by defining a more
general functional integral,
\beq
\label{fancyzsource}
{\cal Z}\left[j\right] ~\equiv~
\int d\nu_f~ d\nu_i~ D\nu ~
e^{B_f[0, \nu_f] ~+~ B_i[0,\nu_i]  ~+~
iS_2\,\left[\,\nu\,\right] ~-~ \int d^3k~ j(k)\,\nu_i(-k)} ~,
\eeq
in terms of which (\ref{fancyz}) may be written
\beq
\label{fancyz2}
{\cal Z}\left[\phi^c\right] ~=~
\left\{ 1 ~+~ \alpha_R\cdot{\delta\over\delta j} \right\} ~
\left\{ 1 ~+~ \alpha_L\cdot{\delta\over\delta j}\right\}~
{\cal Z}\left[j\right] ~\rule[-3mm]{0.2mm}{6mm}_{\, j=j_*} ,
\eeq
where we will set the arbitrary current $j(k)$ equal to
\beq
\label{current}
j_*(k) ~=~ \sqrt{2\,\omega_{\bf k}}~
\left(\,\alpha_R(k) ~+~ \alpha_L(k)\,\right)~
e^{-i\omega_{\bf k}\, T_i} ~,
\eeq
after the functional differentiation.
The stationary point of (\ref{fancyzsource}) is a ``classical''
fluctuation field $\nu^c$ satisfying
\beq
\label{flucteqn}
\Delta\left[ x\right]~ \nu^c\left(x\right) ~=~ 0 ~,
\eeq
with boundary conditions
\beq
\label{fluctbcs}
i\,\dot{\nu}^c_i(k) ~+~ \omega_{\bf k}\nu^c_i(k) ~=~ j(k)
\hspace{0.5cm}
{\rm and}\hspace{0.5cm}
-i\,\dot{\nu}^c_f(k) ~+~ \omega_{\bf k}\nu^c_f(k) ~=~ 0  ~.
\eeq
The solution of this linear homogeneous boundary value problem
can be expressed  in terms of a Greens function
\beq
\label{fluctsoln}
\nu(x) ~=~ \int d^3q ~G\left(\,\vec{x},\vec{q};t\,\right) j(\vec{q})
{}~,
\eeq
where $G$  obeys
\beq
\label{gfeom}
\Delta\left[ x\right]~ G\left(\,\vec{x},\vec{x}^\prime;t\,\right) ~=~
0 ~,
\eeq
with boundary conditions in terms of its Fourier transform
\beq
\label{gfbci}
i\,\dot{G}\left(\,\kvec,\vec{q};T_i\,\right) ~+~
\omega_{\bf k}\, G\left(\,\kvec,\vec{q};T_i\,\right)~=~
\delta^3\left( \kvec - \vec{q} \right)
\eeq
and
\beq
\label{gfbcf}
-i\,\dot{G}\left(\,\kvec,\vec{q};T_f\,\right) ~+~
\omega_{\bf k}\, G\left(\,\kvec,\vec{q};T_f\,\right) ~=~ 0 ~.
\eeq

Using the above expressions, we obtain a compact expression for
(\ref{fancyzsource})
entirely in terms of this Greens function, evaluated at $t = T_i$,
and the known initial state source.
\beq
\label{zresult}
{\cal Z}\left[j\right] ~=~ {\rm det}^{-1/2}\left[\Delta\right]~
\exp~{\left(\, \frac{1}{2}\int d^3k~d^3q~j(\kvec )~
G\left(\kvec,\vec{q};T_i\right)~j(\vec{q}) \,\right)} ~.
\eeq
The exponent appearing here is due entirely to the boundary
functional
$B_i$ in (\ref{fancyzsource}).  All other terms vanish on the
stationary
phase solution (\ref{fluctsoln}).
With this result,
we obtain the Gaussian integral (\ref{fancyz2}) by functional
differentiation and setting $j = j_*$
\beq
\label{zfini}
{\cal Z}\left[\phi^c\right] ~=~ {\rm det}^{-1/2}\left[\Delta\right]~.
\eeq
Several terms in (\ref{fancyz2}) have vanished in the limit of
$T_i\rightarrow\infty$,
most notably initial state factors. This is because both $j_* (k)$
and our
inner
product (\ref{integral}) contain rapidly oscillating factors $e^{- i
\omega_k
T_i}$.
This rapid oscillation allows the application of the Riemann-Lebesgue
lemma
in the limit $T_i\rightarrow -\infty$, which guarantees that for any
function
$f(k)$ whose Fourier
transform exists,
\beq
\label{RL}
\lim_{T_i \rightarrow -\infty}   \int dk~f(k)~e^{- i \omega_k T_i}
{}~=~ 0.
\eeq

Here the functional determinant ${\rm det}^{-1/2}\left[\Delta\right]$
is
actually the product of three determinants: the standard
4-dimensional
determinant of the operator $\Delta$ satisfying Feynman boundary
conditions,
and two 3-dimensional determinants representing the edge fluctuations
$ \nu ( \vec{x}, T_{i,f} )$.
The latter determinants can be expressed in terms of a homogeneous
Greens function similar to  $G$ described above.

We have demonstrated that the kernel of the \mbox{S-matrix}
(\ref{s2})
is obtained to leading order in $g^2$  from the action of the
classical solution and the determinant of quadratic fluctuations
around it.
\beq
\label{result}
S\left[ {\bf b}^*, 2 \right] ~=~ \int d\mu~
e^{B_f[{\bf b}^*, \phi^c_f ] ~+~B_i[0,\phi^c_i] ~+~
iS\,\left[\,\phi^c\,\right]}~{\rm det}^{-1/2}\left[\Delta\right]~
\left[\, 1 + {\cal O}(g^2)\,\right] ~.
\eeq
The classical factors here are known functions, depending only
on the initial and final boundary values of the classical field,
(\ref{phialpha}) and (\ref{tf2}).
Thus, we may further reduce (\ref{result}) to
\beq
\label{result2}
S\left[ {\bf b}^*, 2 \right] ~=~ e^{-1}\int d\mu~
e^{\bar{N}/2 ~+~
iS\,\left[\,\phi^c\,\right]}~{\rm det}^{-1/2}\left[\Delta\right]~
\left[\, 1 + {\cal O}(g^2)\,\right] ~.
\eeq
where $\bar{N} \equiv \int d^3k~ \cohbk b_{\bf k}$ is the average
number
of particles in the final coherent state, which depends implicitly on
the classical field.
The origin of this factor is the use of coherent states
which are not normalized.
The standard normalization for coherent states which we use here is
\beq
\label{norm}
\langle{\bf b}^*\vert{\bf b}^*\rangle = e^{\bar{N}} ~.
\eeq
The correctly normalized  \mbox{S-matrix} element would not
include this factor, leaving the classical action and the
determinant.

There are {\em no} additional tree-level radiative corrections to
(\ref{result2}),
as there are in an expansion around a stationary point which does
not obey the correct boundary conditions (c.f.-- the
instanton~\cite{TIN,KRT}).
This is a direct result of the fact that our classical solution obeys
both the correct equation of motion and the correct boundary
conditions.

It is worth noting that our semiclassical expansion differs from the
usual one in a crucial way.
Our classical background depends on the coupling constant itself.
In the usual instanton calculation,
the field equations and boundary conditions can be made independent
of $g$ by rescaling $\phi \rightarrow \bar{\phi} = g \phi$.
In our case,
since the boundary conditions are fixed by the wave packet shapes
$\alpha_{R,L} (\vec{k})$, the equations of motion for $\phi$
contain a dependence on $g$ which cannot be scaled away.
It is clear that a change in $g$ will lead to a different nonlinear
evolution of $\phi$ and hence a {\it different} classical trajectory.
In the limit of strictly zero coupling constant,
the evolution of the initial wave packet is clearly trivial,
and we get only a contribution to the ``diagonal'' part of the
\mbox{S-matrix} (i.e.-- no scattering).
Therefore, in order to obtain a nontrivial result, we have to take
$g$
large enough to find an interesting classical trajectory, and yet
small
enough to justify the semiclassical $g^2$ expansion derived above.
We have not demonstrated that the two requirements can be satisfied
simultaneously in any theory.
This can only be checked {\it a posteriori} by computing the size of
the subleading  ${\cal O}(g^2)$ corrections in the background
of a proposed trajectory.
Of course, our experience with perturbation theory leads us to
believe
that for $g^2 \ll 1$ the corrections will be small.
Thus, a classical trajectory found using weakly coupled equations
of motion is likely to provide a controlled approximation.

\section{COMPLEX TRAJECTORIES}

In the preceding discussion,
we have focused on classical trajectories dependent on a real time
variable.
These are relevant to classically allowed amplitudes,
such as those for multiparticle scattering or transitions at energies
well
above an energy barrier in configuration space.
As we noted previously,
the existence of a real time trajectory implies an unsuppressed
scattering
amplitude for the corresponding transition.
However,
there are many instances in which classically disallowed transitions
are of interest.
In particular,
baryon number violation in the electroweak theory at energies less
than
the sphaleron energy $E_s \sim M_w/\alpha_w \sim 10~ TeV$
is classically forbidden and therefore cannot be described by the
same type of real time trajectories.
The application of our formalism to such cases therefore requires
some
generalization, which we will describe in this section.

Consider the initial value problem given by (\ref{eom}) and
(\ref{ti}).
Since both the initial field values $\phi_i (x)$
and their time derivatives $\dot{\phi}_i (x)$ are given,
the classical field has a well defined energy.
Since real time trajectories conserve energy,
it is clear that a suitable trajectory will not exist
for some choices of initial and final conditions.
In particular, in the electroweak theory for $E_{cl} < E_s$,
there are no real time trajectories which connect initial
and final conditions with different topological charge.

In order to find such trajectories,
previous authors~\cite{KRT3,Miller,GP} have proposed complex
{\it time} paths,
which involve alternately Minkowskian, Euclidean and
Minkowskian paths joined together at two times, say $0$ and $iT$
as in figure 2.
In this approach,
$t = 0$ and $t = iT$ must be turning points
($\dot{\phi} (\vec{x}, T_i) = 0$ for all $\vec{x}$)
of the field equations with given initial conditions.
As is familiar from one dimensional quantum mechanics~\cite{Miller},
complex time contours describe an incident configuration
which reaches a turning point,
and then tunnels under the barrier in Euclidean time,
returning to real time at the second turning point
upon reaching the other side of the barrier.
The field $\phi(x)$ can be taken to be real on the entire contour
because of the existence of the two turning points~\footnote{A fact
guaranteed in one dimensional QM, but not in higher dimensions or
in field theory ~\cite{GP}.}.
The action is then purely imaginary on the Euclidean part of
the contour and yields the familiar exponential tunneling
suppression.
\beq
\label{suppression}
\vert\,\langle f\,\vert\, i\,\rangle\,\vert^2~\sim~e^{-2\,{\rm
Im}\,S} ~.
\eeq

As a matter of practice,
one might try to find a suitable complex time trajectory by
integrating
our initial conditions (\ref{u}) forward in  real time in hopes that
a turning point is encountered.
Then, the time variable should be Wick rotated $t \rightarrow ix_4$
and the turning point configuration integrated forward in Euclidean
time until another turning point is encountered.
Finally,
the contour should be rotated back to real time and the second
turning point integrated forward until it achieves its asymptotic
plane
wave state.  The final state $\vert b^*\rangle$
can be read off from (\ref{tf2}).

$$
\beginpicture
\setplotarea x from -65 to 65 , y from -40 to 55
\putrule from 0 -20 to 0 100
\arrow < 4 pt> [0.4,1.5] from 0 100  to 0 103
\put{${\rm Im}~t$} at 0 120
\putrule from -80 20 to 80 20
\arrow < 4 pt> [0.4,1.5] from 80 20 to 83 20
\put{${\rm Re}~t$} at 100 20
\putrule from -80 60 to 0 60
\arrow < 4 pt> [0.4,1.5] from  -40 60  to -37 60
\arrow < 4 pt> [0.4,1.5] from   40 20  to  43 20
\put{$i\, T$} at 15 60
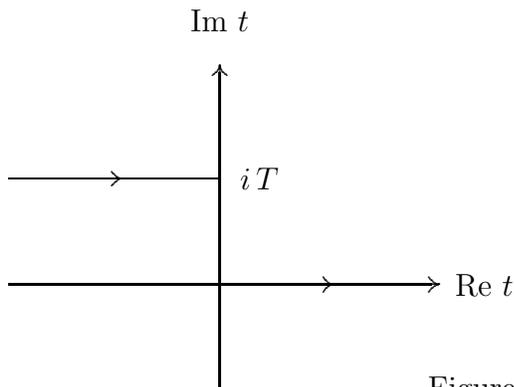
\put{ Figure 2: Complex time contour} at 160 -20
\endpicture
$$

However, in the most general case,
there is no guarantee of encountering turning points.
Here we propose a more general method for extremizing S-matrix
elements
for which there is not a corresponding real, Minkowskian trajectory.
Rather than complexifying time,
we instead search for a general {\it complex} saddlepoint
configuration
satisfying (\ref{eom}) and (\ref{ti}).
Our motivation stems from the usual method of steepest descents
applied
to a real integral of the form
\beq
\label{I}
I(h) ~=~ \int_{x_a}^{x_b} ~dx~ e^{\frac{1}{h} f(x)}.
\eeq

In general, the asymptotic series in $h$ for this integral can be
obtained
by expanding about a saddlepoint of the function $f(x): f'(x)|_{x^*}
= 0$.
However, the saddlepoint $x^*$ is often complex, and to apply the
method,
one must first deform the original real integral into the complex
plane
(see figure 3).

In our case,
we first think of the Feynman Path Integral as the product of a large
number of regular integrals by discretizing spacetime:
\beq
\int D\phi ~e^{iS[\phi]} ~~  \rightarrow ~~  \int  \prod_{x} ~ d
\Phi_{x} ~
e^{ i S( \Phi_{x},   \Phi_{x \pm 1} )}     .
\eeq
For a real field $\phi(x)$ each integral in the product is a real
integral
like (\ref{I}).
We generalize the method of steepest descents by allowing the
variable of
each integration, $\Phi_{x}$, to become complex.
There should be no obstruction to this generalization as
the function $S( \Phi_{x},   \Phi_{x \pm 1} )$ is analytic in the
variables $\Phi_x$.
This procedure leads us to search for a complex trajectory $\{ \Phi_x
\}$
(or in the continuum, $\phi^c (x)$) satisfying (\ref{eom}),
(\ref{ti}) and
(\ref{tf}).
The integral over small fluctuations about $\phi^c(x)$ must be
performed
along the path of steepest descent in configuration space.
In other words, along small $\nu(x)$ fluctuations which keep the
imaginary
part of $iS\left[ \phi^c (x) + \nu(x)\right]$ constant.

$$
\beginpicture
\setplotarea x from -65 to 65 , y from -40 to 55
\putrule from 0 -20 to 0 110      \put{${\rm Im}~x$} at 0 140
\arrow < 4 pt> [0.4,1.5] from 0 100  to 0 130
\putrule from -75 20 to 130 20    \put{${\rm Re}~x$} at 160 20
\arrow < 4 pt> [0.4,1.5] from 130 20 to 133 20
\circulararc 90 degrees  from  -60 20 center at -60 30
\circulararc -70 degrees  from  -50 30 center at 0 30
\setlinear
\plot
  -17  77
   43  95
/
\circulararc -90 degrees  from 43 95 center at 50 70
\plot
  75  77
  85  33
/
\circulararc -90 degrees from 102 20 center at 100 35
\put{$x_b$} at 102 10       \put{$\times$} at 33 92
\put{$x^*$} at 35 78        \put{$x_a$} at -60 10
\arrow < 4 pt> [0.4,1.5] from -7 80 to -4 81
\arrow < 4 pt> [0.4,1.5] from 79 59 to 81 51
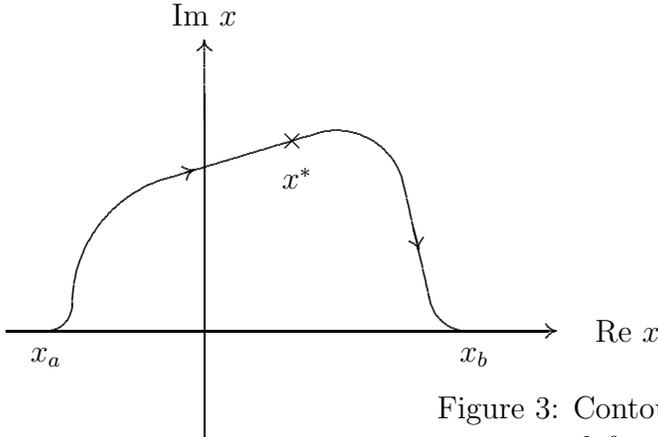
\put{ Figure 3: Contour in complex plane,} at 180 -10
\put{deformed through saddlepoint $x^*$} at 225 -25
\endpicture
$$

For a complex field,
there is no longer a relation between the positive and negative
energy
Fourier components.
Without loss of generality, we can write
\beq
\label{phicomplex}
\phi_i(\vec{p}) ~=~ {1\over \sqrt{2\omega_{\bf p}}}~
\left(\,
u_{\bf p}~
e^{-i\omega_p T_i} ~+~  v_{\bf p}~e^{i\omega_p T_i}
\,\right) ~,
\eeq
where $u_{\bf p}~,~ v_{\bf p}$ are independent complex functions.
The initial boundary condition (\ref{ti}) is satisfied by
\beq
\label{ucomplex}
u_{\bf p} ~=~
{\alpha_R(\vec{p})\over\int d^3k~\alpha_R(\vec{k})~v_{\bf k}}~+~
{\alpha_L(\vec{p})\over\int d^3k~\alpha_L(\vec{k})~v_{\bf k}}~,
\eeq
where $v_{\bf k}$ is arbitrary.
The initial  configurations which satisfy
(\ref{phicomplex}), (\ref{ucomplex}) have negative frequency modes
which ``resemble'' the wave packet $\alpha_L + \alpha_R$ (up to
  complex multiplicative factors) and positive frequency modes which
are arbitrary.

It is then clear that there are an infinite number of complex
trajectories
which can result from evolving (\ref{phicomplex}) using (\ref{eom}).
The final quantum state $\vert {\bf b}^* \rangle$
which results from a given trajectory depends only on the positive
frequency
component of the trajectory in the far future (see Eq. (\ref{tf})).
The multiplicity of trajectories  allows a very large set of
scattering amplitudes to be addressed within our formalism.
In fact,
it seems in general there always exists  a complex trajectory
which satisfies (\ref{eom}), (\ref{ti}) and (\ref{tf}) simultaneously
for any $\alpha(k)$ and ${\bf b}^*$.
This is because the initial and final conditions (\ref{ti}) and
(\ref{tf})
each correspond to conditions on a combination of
$\phi (x)\vert_{T_i}$ and  $\dot{\phi}(x)\vert_{T_i}$,
and $\phi (x)\vert_{T_f}$ and $\dot{\phi}(x)\vert_{T_f}$,
respectively (mixed Dirichlet-Neumann boundary conditions).
Since the equations of motion (\ref{eom}) are second order in time,
this mixed set of boundary conditions should be enough to specify
a unique solution.

Since the trajectories found here are complex, the
action $i S[ \phi^c]$ may contain a real suppression factor.
This is to be expected, as some of the trajectories will correspond
to
classically disallowed tunnelling transitions.
Thus, although potentially {\em any} S-matrix element can be
approximated in
this way, the result for particular trajectories may be exponentially
small.

\section{RELATION TO PREVIOUS WORK}

Here
we comment on the relationship between our approach and
some previous work by other authors on high energy
multi-particle and anomalous processes  in electroweak theory.
First,
we will recall the expansion around the constrained instanton in
electroweak theory~\cite{TIN,KRT}.
This approach neglects effects of initial and final states on the
saddlepoint and results in a low energy ($E\ll E_*\simeq
M_w/\alpha_w$)
approximation to the total two-particle cross-section.
Then, we remark on a previous approach to account for the impact
of initial and final states on the saddlepoint~\cite{MMY}
and comment on the formal resemblance which it bears to our approach.

A few of the properties of the electroweak instanton will make clear
the relationship to our work:
({\em i})
The electroweak instanton is a solution of the Euclidean equations
of motion.  So, it will be necessary to rotate our real time
formalism
to Euclidean space in the  analysis above, $ t\rightarrow ix_4 $.
We would then be considering a saddlepoint approximation to an
on-shell
truncated Euclidean Greens function.
({\em ii})
The electroweak instanton has finite Euclidean action, $ S =
8\pi^2/g^2 $,
and therefore satisfies {\em vacuum} boundary conditions.
So, it is a suitable saddlepoint point for a Greens function with
external fields only insofar as the effect of initial and final
states
on the saddlepoint are neglected (dropping the right hand side
of (\ref{ti}) and (\ref{tf})).

Since the instanton does not obey the correct boundary
conditions to be a saddlepoint of a scattering amplitude~\cite{TIN},
linear terms in the fluctuation expansion do not cancel,
and the expansion of the scattering amplitude entails corrections
which are formally large $ {O \cal }(1/g^2) $  in the exponent.
Fortunately,
these corrections are under control and calculable for sufficiently
small
energies, $ E\ll E_* $ ~\cite{KRT}.
The effect of the final states can be taken into account,
in a perturbative expansion in powers of $ x\equiv E/E_*\ll 1
$~\cite{KRT}.
For instance,
the total inclusive cross section in the one-instanton sector is
given
by the well-known result
\beq
\label{holygrail}
\sigma_{tot} (E) ~\sim~
\exp \left\{  {16\,\pi^2\over g^2}
\left( -1 + c_1 x^{4/3} + c_2 x^2 +
{\cal O}\left(x^{8/3}\right)\right)\right\} ~,
\eeq
in the limit $E/E_* = fixed$ and $g\rightarrow 0$.
Here $\sim$ implies that only the exponential behavior
of the cross-section is shown.
The first term here is just twice the instanton action,
the 'tHooft suppression for vacuum tunnelling,
while the next term indicates the exponential growth of the
cross-section
at low energies~\cite{BNV}.
The higher order terms are determined from all tree-level corrections
to the many soft ($E\simeq M_w$) final state particles,
and the first few coefficients $c_i$ are known.
Thus, provided the energy is sufficiently low $ E \ll E_* $,
tree-level corrections to the final state can be described
{\it semiclassically}.

It has not been similarly demonstrated that the effect of
{\it initial} state corrections, involving many loops,
can be described semiclassically.
However,
some results indicate that these corrections may run counter to naive
intuition and the initial two particle state may be described
semiclassically as well~\cite{IS,MMY,TIN}.
As we noted in Section 4,
initial state corrections in the instanton expansion
appear as factors of the residue of the propagator in the
instanton background.
The residue of the propagator has hard high energy behavior
( $\sim g^2 s$ ) where $s$ is the center of mass energy~\cite{IS},
so that initial state corrections can contribute at the
{\it semiclassical}  level when $s$ is of order $1/g^4$.
Mueller~\cite{IS} has shown that they contribute first at order
$\left(E/E_*\right)^{10/3}$ to (\ref{holygrail}).

By contrast, in our formalism
the classical background field is constructed to solve the correct
multi-particle boundary value problem derived in Section 2.
So, there are no additional tree-level corrections in the limit
$E=fixed$ while $g\rightarrow 0$.
All corrections are formally suppressed by powers of $ g $,
as shown in Section 4.
The size of subleading corrections to our result depend on the
properties of the Feynman propagator in the background of our
proposed classical trajectory, $ \phi^c $.
It would therefore be very interesting to understand the high energy
behavior of the propagator in our background field.
Also, for the purposes of direct comparison to (\ref{holygrail}),
it would be interesting to obtain its behavior in the same limit,
where $E/E_* = fixed$ while $g\rightarrow 0$.
We have reason to expect that the propagator behaves less severely
than the instanton propagator in this limit,
due to the lack of translational zero modes of the background
field~\cite{MMY}.

A previous approach to account for the impact of initial and final
states on the saddlepoint~\cite{MMY} bears a resemblance to ours but
results in a different classical equation.
We discuss it here for the purposes of comparison.
These authors extremized Minkowskian $n$-point Greens functions of
electroweak gauge fields
\beq
\label{MMY1}
G^{(n)}\left(p_1, p_2, \ldots\right) ~=~
\int {\cal D}A ~A^{\mu_1a_1}(p_1) \cdots A^{\mu_n a_n}(p_n)~
e^{iS\left[A\right]/g^2 } ~.
\eeq
It is not clear to us  whether the Greens function is the proper
quantity to extremize to yield information about the
corresponding scattering amplitude (and ultimately the
cross-section).
It may be that extremizing the entire Greens function,
rather than its LSZ projection, is too strict a requirement.

An equation of motion for the saddlepoint of (\ref{MMY1}) can be
derived
in a manner with formal similarity to our own.
To do so, first express the $n$ fields as
an exponential, using $A = \exp\ln A $,
and then extremize the exponent.  This yields~\cite{MMY}
\beq
\label{MMY2}
{\delta S\over \delta A^{\nu b}(x)} ~=~
i~\sum_i~\delta^{\mu_i \nu}~\delta^{a_i b}~
{e^{ip_i\cdot x}\over A^{\mu_i a_i}(p_i)} ~.
\eeq
This is a non-linear integro-differential  field equation,
depending on both the field $A(x)$ and its Fourier transform $A(p)$.
Not much is known about equations of this type, and despite
much effort (\ref{MMY2}) defies solution~\cite{MCL}.
It appears more difficult to solve in practice than the boundary
value
problem we have derived (\ref{eom}), (\ref{phialpha}) and (\ref{tf}).

Before concluding I would like to point out a difference between
studying scattering via the Minkowskian and Euclidean functional
integral.
The Euclidean counterparts to our solutions (i.e. their analytic
continuations
to imaginary time $t \rightarrow ix_4$) are very badly behaved
asymptotically.
That is, the `wrong frequency' components required by the prescence
of
particles
in the initial and final states (see equations (\ref{ti}) and
(\ref{tf}) )
imply an exponential blow-up of the solution
at large imaginary times. In other words, our BC's are incompatible
with
the usual Feynman BC's (only positive frequencies in the future, and
negative
in the past)
which arise automatically from requiring finite action in Euclidean
space.

\section{SUMMARY}

In this lecture I
have outlined a connection between certain classical solutions and
corresponding quantum S-matrix elements in a weakly coupled field
theory.
The connection is made via a semiclassical approximation which
appears to
be controlled at small coupling,
regardless of the scattering energy.
This is in contrast to methods which involve expansions about
instanton-like
trajectories (i.e.-- satisfying vacuum boundary conditions),
which have been shown to break down at nonperturbative
energy scales $E \sim m/g^2$.
Real Minkowskian trajectories can be found by the straightforward
time
integration of a
well-defined initial value problem which is determined by the
inital quantum state.
Complex Minkowskian trajectories require the implementation
of mixed boundary conditions at both initial and final times.
This procedure, while more difficult in practice, can be guaranteed
in
principle to describe any wave packet to coherent state transition
specified
by the complex functions $\alpha(k), b^*_k$.
The formalism described here can thus be used to compute
intrinsically nonperturbative scattering amplitudes in
a variety of field theories.

Some examples discussed here are the problem of baryon number
violation in the
electroweak theory and multiparticle scattering in gauge and scalar
theories.
The discovery of nontrivial or ``interesting'' real classical
trajectories in
these theories can now be related to unsuppressed scattering
amplitues which are potentially observable experimentally.
Alternatively, nontrivial complex trajectories, of which there are
an infinite number,
allow the calculation of a much larger class of scattering
amplitudes,
including some that are classically forbidden and therefore involve
tunneling.
We hope that our results will stimulate future numerical work in this
area,
particularly the search for nontrivial trajectories.

\vskip 1.0cm
\centerline{\bf Acknowledgements}
This talk is based on research \cite{GHP} performed in collaboration
with
Thomas Gould and Erich Poppitz, both of
The Johns Hopkins University.
SDH acknowledges support from the National Science Foundation
under grant \mbox{NSF-PHY-87-14654,}
the state of Texas under grant \mbox{TNRLC-RGFY-106,}
the Harvard Society of Fellows and an SSC Fellowship.

\nc{\ib}[3]{        {\em ibid. }#1:#3 (19#2) }
\nc{\np}[3]{        {\em Nucl. Phys. }#1:#3 (19#2) }
\nc{\pl}[3]{        {\em Phys. Lett. }#1:#3 (19#2) }
\nc{\pr}[3]{        {\em Phys. Rev.  }#1:#3 (19#2) }
\nc{\prep}[3]{      {\em Phys. Rep.  }#1:#3 (19#2) }
\nc{\prl}[3]{       {\em Phys. Rev. Lett. }#1:#3 (19#2) }

\end{document}